\documentclass[aip,reprint,amsfonts,amsmath,amssymb,citeautoscript,floatfix]{revtex4-1}
\usepackage{bm}
\usepackage{dcolumn}
\usepackage{graphicx}
\usepackage{parskip}
\usepackage{enumerate}

\usepackage{upgreek}
\usepackage{psfrag}
\usepackage{color}
\definecolor{amber}{rgb}{1.0, 0.75, 0.0}
\definecolor{red}{rgb}{0., 0., 0.}
\setcitestyle{numbers,square}

\begin{document}

\title{dense packings of hard circular arcs}

\author{Juan Pedro Ram\'irez Gonz\'alez}
\affiliation{
Departamento de F\'isica Te\'orica de la Materia Condensada, \\
Universidad Aut\'onoma de Madrid,\\
Ciudad Universitaria de Cantoblanco, 
E--28049 Madrid, Spain
}

\author{Giorgio Cinacchi}
\affiliation{
Departamento de F\'isica Te\'orica de la Materia Condensada,  \\
Instituto de F\'isica  de la Materia Condensada (IFIMAC),    \\
Instituto de Ciencias de Materiales ``Nicol\'{a}s Cabrera'', \\
Universidad Aut\'onoma de Madrid,\\
Ciudad Universitaria de Cantoblanco, 
E--28049 Madrid, Spain
}

\date{\today}

\begin{abstract}
This work investigates dense packings of 
congruent hard infinitesimally--thin circular arcs in 
the two--dimensional Euclidean space.
It focuses on 
those denotable as major 
whose subtended angle $\uptheta \in \left ( \uppi, 2\uppi \right ]$. 
Differently than 
those denotable as minor 
whose subtended angle $\uptheta \in \left [0, \uppi \right]$, 
it is impossible for 
two hard infinitesimally--thin circular arcs with $\uptheta \in \left ( \uppi, 2\uppi \right ]$
to arbitrarily closely approach 
once they are arranged in 
a configuration, e.g. on top of one another, 
replicable ad infinitum without introducing any overlap.
This makes these hard concave particles,
in spite of being infinitesimally thin, 
most densely pack with 
a finite number density. 
This raises the question as to
what are these densest packings and 
what is the number density that they achieve.
Supported by Monte Carlo numerical simulations, 
this work shows that 
one can analytically construct 
compact closed circular groups of 
hard major circular arcs in which 
a specific, $\uptheta$--dependent, number of them 
(anti--)clockwise intertwine.
These compact closed circular groups then arrange on a triangular lattice.
These analytically constructed densest--known packings are compared to 
corresponding results of Monte Carlo numerical simulations 
to assess whether 
they can spontaneously turn up.
\end{abstract}
\maketitle

\section{introduction}
\label{introduction}
Systems of hard 
(i.e. non--interacting except for non--intersecting)
particles are 
elementary model systems with which 
to investigate (condensed) states of matter \cite{solids,liquids}. 
Out of the many aspects of this investigation 
{\color{red}
\cite{allenfrenkel,torquato1,mvmr,dijkstra,escobedo,torquato2}}, 
one is the determination of 
those packings 
(i.e. single configurations) that 
exhibit the maximal density \cite{torquato1,torquato2}.

In the two--dimensional Euclidean space (${\mathbb{R}}^2$),
{\color{red} it has long been mathematically proven 
that hard circles, the simplest hard particles, 
most densely pack in a triangular lattice 
\cite{fejes}}.
Equally mathematically proven {\color{red}has been}  
that hard convex particles, 
if centro--symmetric, most densely pack in a lattice \cite{fejes2} while, 
if non--centro--symmetric, pack in a double lattice that 
covers, as a minimum, a fraction of ${\mathbb{R}}^2$ equal to 
$\displaystyle \frac{\sqrt{3}}{2}$ \cite{kuperberg}.  
For hard concave particles,  
the sole conjecturing what may be the densest packings may not be immediate,   
leaving aside devising a convincing mathematical proof that these packings are such indeed.
In this context, numerical methods, that originated in physics, 
such as the Monte Carlo and molecular dynamics methods \cite{MCorigin,MDorigin,MCNPT,parrinello,MCallen}, 
become increasingly useful.
{\color{red} In two recent works, event--driven molecular dynamics was used 
to determine the densest--known packings of hard convex and concave superdiscs \cite{superdiscs} while 
a method of the Monte Carlo} type \cite{ASC} was used 
to determine the densest--known packings of 
an ample variety of hard convex and concave particles \cite{atkinson}.

\begin{figure}
\centering
\includegraphics[scale=0.75]{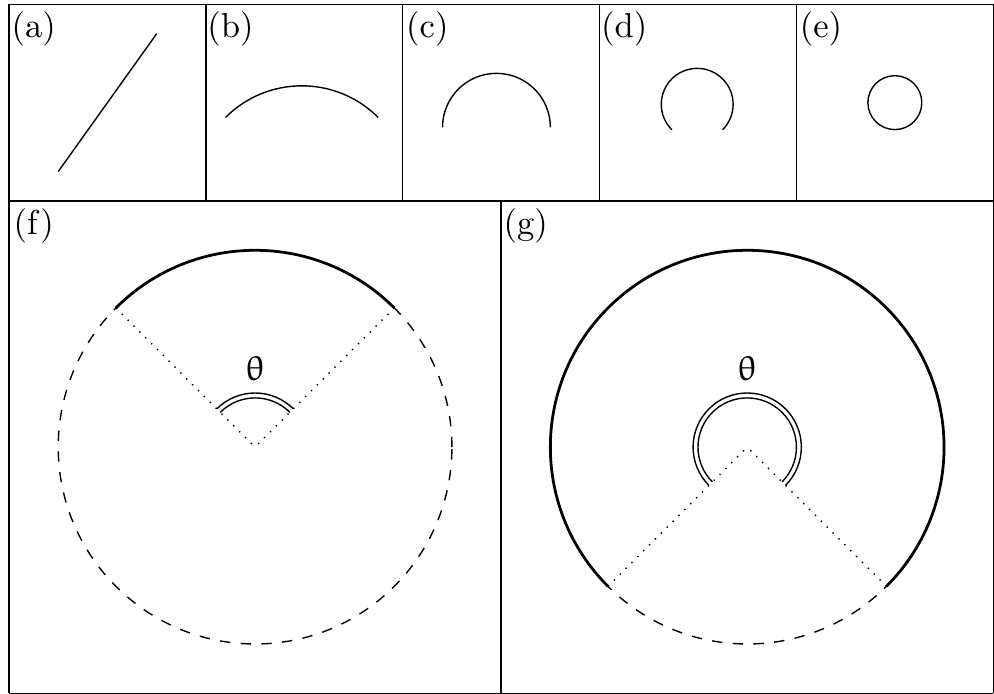}
\caption{
Examples of circular arcs with different subtended angle $\uptheta$ and equal length: 
(a) $\uptheta=0$; (b) $\uptheta=\uppi/2$; (c) $\uptheta = \uppi$; (d) $\uptheta=3\uppi/2$; (e) $\uptheta=2\uppi$. 
{\color{red}Examples of a minor circular arc with $\uptheta \le \uppi$ (f) and of a major circular arc with $\uptheta > \uppi$ (g):
these circular arcs are drawn with a continuous line while the circumferences of their parent circles are drawn with a discontinuous line;
the subtended angle $\uptheta$ is explicitly indicated.} 
}
\label{figura1p}
\end{figure}

Out of the many generalisations of a circle, 
one views it as a special circular arc.
It is then natural to consider, as hard, generally non--circular, particles, 
hard circular arcs in the entire interval of the subtended angle $\uptheta$,
from the linear segment limit, corresponding to $\uptheta=0$,
to the circle, corresponding to $\uptheta=2\uppi$ [Fig. \ref{figura1p} (a--e)].
These hard particles are deceptively simple:
they are {\color{red}generally} concave and, in spite of being infinitesimally thin,
they generally exclude an area to one another.
The class of circular arcs can be divided into two sub--classes:
(\textsc{i})  those circular arcs with $\uptheta \in [0,\uppi]$,
i.e. minor circular arcs [Fig. \ref{figura1p} (f)], from the linear segment limit to the semi--circular arc;
(\textsc{ii}) those circular arcs with $\uptheta \in (\pi, 2\uppi]$,
i.e. major circular arcs [Fig. \ref{figura1p} (g)], from the semi--circular arc to the circle.
Two hard particles of the former sub--class can arbitrarily closely approach:
e.g. once they are arranged on top of one another.
Since this arrangement is replicable ad infinitum, without introducing any overlap,
there exist infinitely dense packings of these hard particles.
For two hard particles of the latter sub--class,
this two joint conditions do not hold any more.
It is then natural to inquire  
what are their densest packings and 
what is the number density that they achieve.

This work attempts to answer this question.
Supported by Monte Carlo numerical simulations,
one can 
peculiarly interlace a specific number, depending on $\uptheta$, of 
hard major circular arcs to analytically construct compact closed circular groups;
{\color{red}one can then arrange these compact closed circular groups 
on a triangular lattice}.

To appreciate the mechanism for constructing these compact closed circular groups, 
it is useful to examine the characteristics of the excluded area of 
two hard (major) circular arcs (section \ref{excar}).
Once this has been accomplished, 
the mechanism for constructing compact closed circular groups of 
a specific number of them is devised{\color{red}; it  
immediately leads to the densest--known packings by 
arranging these compact closed circular groups} on a triangular lattice (section \ref{condenspack}).
To asses whether these compact closed circular groups can spontaneously 
either make up or undo,
specific Monte Carlo calculations are then carried out.
In general, the closed circular groups that do form on compression in these Monte Carlo calculations 
have a number of constituent hard major circular arcs smaller than 
that achievable by analytic construction;
the analytically constructed compact closed circular groups are nevertheless able to 
unfasten on decompression in these Monte Carlo calculations (section \ref{MC}).
This suggests that
the spontaneous replication of  
the complete mechanism responsible for 
the formation of these optimal packings of hard major circular arcs, 
although possible in theory, 
will be extremely improbable in practice,
but
similar sub--optimal packings of hard {\color{red}(colloidal, granular)} major circular arcs 
will nevertheless be readily accessible
(section \ref{conclusion}).

\section{excluded area and construction of the densest--known packings}
\subsection{excluded area}
\label{excar}

In a two--dimensional space, 
the excluded area of two hard particles is 
that surface whose constituent points  
the centroid of one particle cannot occupy  
due to the presence of the other particle 
otherwise the two hard particles would overlap.
For two hard circular particles, 
the excluded area is the area of a circle 
whose centre coincides with the centre of one particle and 
whose radius is the sum of the radii of the two particles. 
For two hard non--circular particles,  
the excluded area depends on their fixed relative orientation.
One of them, particle \texttt{1}, can be held fixed, 
e.g. with its centroid at the origin of the Cartesian reference frame and 
an axis along one of the two Cartesian axes, 
while the other, particle \texttt{2}, 
whose axis is rotated by a certain angle $\uppsi$ with respect to the particle \texttt{1} axis, 
displaces around, freely except for the constraint that it cannot overlap particle \texttt{1}.
While displacing, the particle \texttt{2} centroid is as if it effectively generate 
a region delimited by a boundary{\color{red};} 
all internal points are prohibited positions for the particle \texttt{2} centroid, 
since, if it occupied one of them, particle \texttt{2} would overlap particle \texttt{1},
while all external points are permitted positions for the particle \texttt{2} centroid,
since, if it occupies one of them, particle \texttt{2} does not overlap particle \texttt{1}.  

Their concavity and infinitesimal thinness make 
the area that two hard circular arcs exclude to one another peculiar:
it deserves an examination.

\begin{figure}
\centering
\includegraphics[scale=0.75]{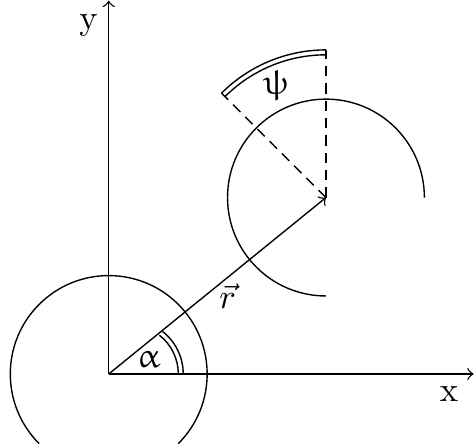}
\caption{
Two circular arcs in ${\mathbb{R}}^2$ with 
the geometric elements that define their pair configuration in 
a (x,y) Cartesian reference frame:
$\uppsi$, the angle that the two symmetry axes form; 
$\vec r$, the vector whose modulus is $r = \left | {\vec r} \right |$ that joins the centres of the two parent circles; 
$\upalpha$, the angle that this vector forms with the x--axis of the Cartesian reference frame. 
}
\label{figura2}
\end{figure}
Since a circular arc is axis--symmetric,
it suffices to consider the angle $\uppsi \in [0,\uppi]$.
Once $\uppsi$ has been fixed, 
a pair configuration of hard circular arcs can be completely defined by $\vec r$, 
the distance vector separating the centres of the parent circles, whose radius is R,
with 
$r = \left | {\vec r} \right |$ the modulus of this vector, and
$\upalpha$ the angle that it forms with the x--axis of the Cartesian reference frame (Fig. \ref{figura2}).
One can define a function 
$\mathsf{f} \left( {\vec r} \,;\, \uppsi \right)$
that takes on 
the value 1 if the two hard circular arcs overlap and the value -1 if they do not,
the overlap condition being established 
according to an exact and suitable overlap criterion.
The excluded--area boundary points can be considered as 
the zeros of $\mathsf{f} \left( {\vec r} \,;\, \uppsi \right)$.

Without much loss of generality,
one can focus on two hard major circular arcs.
In this case, it is very useful to introduce the angle{\color{red}\[
\displaystyle \updelta = \left( \uptheta - \uppi \right) \in (0,\uppi] \,.
\]}For each value of $\updelta$, 
one should distinguish between 
the case $ 0 \le \uppsi \le \updelta $ and the case $ \updelta < \uppsi \le \uppi $.

\begin{figure*}
\centering
\includegraphics[trim=2.0cm 15.0cm 1.0cm 4.0cm, clip, scale=1]{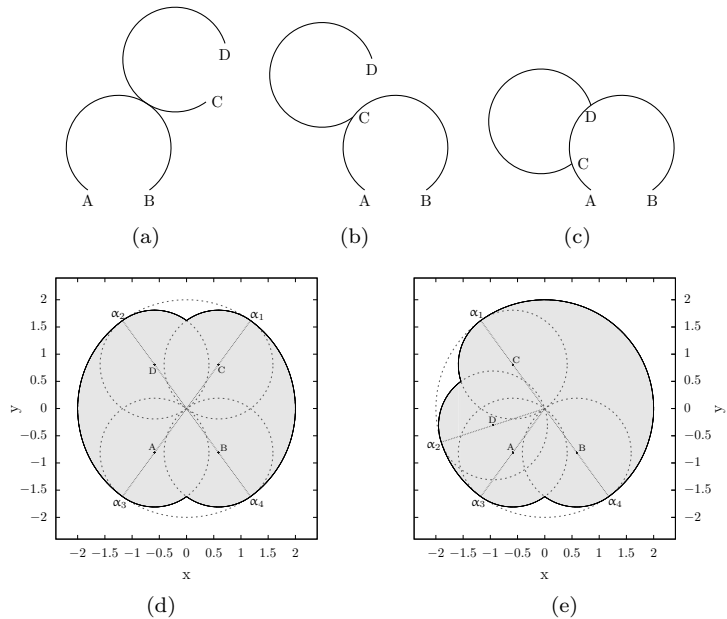}
\caption{
{\color{red} Examples of three} types of contact configuration between two hard circular arcs:
(a) internal point -- internal point; (b) internal point -- extremal point; 
(c) each extremal point of one hard circular arc respectively touches an internal point of the other hard circular arc.
In any of these panels, the extremal points of the hard circular arc held fixed are labelled as A and B while
those of the displacing hard circular arc as C and D. 
The excluded area, the {\color{red}shaded} region enclosed by the black continuous line, of 
two hard major circular arcs with $\uptheta=1.6\uppi$, i.e $\updelta = 0.6 \uppi$,
for the angle of relative orientation $\uppsi = 0$ (d) and $\uppsi =  0.4 \uppi$ (e).  
Discontinuous round lines either correspond to 
the circumference with centre (0,0) and radius 2R or 
to circumferences with the centres that are labelled as the extremal points in panels (a), (b) and (c) and radius R.
The special angles are: 
$\displaystyle \upalpha_1 = \frac{\updelta}{2} + \uppsi$; $\displaystyle \upalpha_2 = \uppi - \frac{\updelta}{2} + \uppsi$;
$\displaystyle \upalpha_3 = \uppi + \frac{\updelta}{2} $; $\displaystyle \upalpha_4 = 2\uppi - \frac{\updelta}{2}$. 
Lengths in panels (d) and (e) are in units of R.
}
\label{figura33}
\end{figure*}
\begin{figure}
\centering
\includegraphics[scale=0.75]{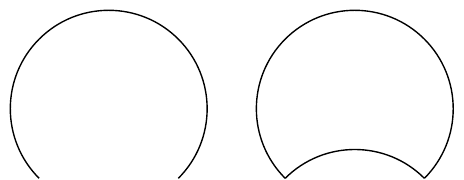}
\caption{
Comparison between a major circular arc that subtends an angle $\uptheta$ (left) and 
a lunate particle formed by 
the juxtaposition of this major circular arc with 
the minor circular arc that subtends the explementary angle $2\uppi - \uptheta$ (right). 
Provided that the angle of their relative orientation is $0 \le \uppsi \le \updelta = (\uptheta - \uppi)$,
there is no difference between 
the excluded area of two hard major circular arcs and 
that of two hard lunate particles.
}
\label{figura55}
\end{figure}

In the case $0 \le \uppsi \le \updelta $, 
one can distinguish between two situations:
either the {\color{red} two respective contact} points at which the two hard circular arcs touch are both internal
{\color{red} [Fig. \ref{figura33} (a)]} or 
{\color{red} at least} one of these contact points is extremal {\color{red} [Fig. \ref{figura33} (b, c)]}.
In the former situation, 
the two hard major circular arcs  behave as two hard circles, 
with the boundary points that follow a circumference of radius 2R [Fig. \ref{figura33} (a)].
In the latter situation, the two hard major circular arcs behave distinctively, 
with the boundary points that follow a circumference of radius R [Fig. \ref{figura33} (b, c)].  
This latter situation occurs for 
$\upalpha \in (\upalpha_1 ,\upalpha_2)$ and $\upalpha \in (\upalpha_3 , \upalpha_4)$ 
with $ \displaystyle \upalpha_1 = \frac{\updelta}{2} + \uppsi$, $ \displaystyle \upalpha_2 = \uppi - \frac{\updelta}{2} + \uppsi$,
$\displaystyle \upalpha_3 = \uppi + \frac{\updelta}{2} $ and $ \displaystyle \upalpha_4 = 2\uppi - \frac{\updelta}{2}$
[Fig. \ref{figura33} (d, e)].
For values of $\upalpha$ equal to 
the mid--values 
$\displaystyle \upalpha_{12} = \frac{\upalpha_1 + \upalpha_2}{2}$ and $\displaystyle \upalpha_{34} = \frac{\upalpha_3 + \upalpha_4}{2}$,
the two hard major circular arcs touch as peculiarly as in Fig. \ref{figura33} (c):
each extremal point of one of them respectively touches an internal point of the other.
On increasing the angle $\uppsi$, 
both extremes of the interval $(\upalpha_1 , \upalpha_2)$  increase of the same quantity $\uppsi$ 
with respect to the values that they have at $\uppsi = 0$ 
[cf. panel (d) with panel (e) of Fig. \ref{figura33}].
While the shape of the boundary changes, the area that it encloses does not: 
the value of the excluded area stays constant throughout the interval $0 \le \uppsi \le \updelta$
[cf. panel (d) and panel (e) of Fig. \ref{figura33}: 
the difference between the area of a circle of radius 2R and the excluded area consists of
the sum of the two arched triangular regions comprised 
between $\upalpha_1$ and $\upalpha_2$ and between $\upalpha_3$ and $\upalpha_4$: 
on going from panel (d), corresponding to $\uppsi=0$, to panel (e),
corresponding to $\uppsi = 0.4 \uppi$, 
the former of these two arched triangular regions rotates {\color{red}round an axis passing through
the point (0,0) and perpendicular to the plane} of 
an angle equal to $0.4 \uppi$ while the latter of these two arched triangular regions does not change: 
hence, that difference does not vary and 
consequently neither does the excluded area].
It is pertinent to observe that, for $0 \le \uppsi \le \updelta$, 
the excluded area of two hard major circular arcs coincides with that 
that two hard lunate particles, 
formed by the juxtaposition of the same major circular arc with the minor circular arc that subtends the explementary angle
(Fig. \ref{figura55}), exclude to one another:
for $0 \le \uppsi \le \updelta$, 
the infinitesimal thinness of two hard major circular arcs and 
the ensuing capability of them of intertwining
do not manifest yet.

\begin{figure*}
\centering
\includegraphics[trim=2.0cm 15.5cm 1.0cm 4.0cm, clip, scale=1]{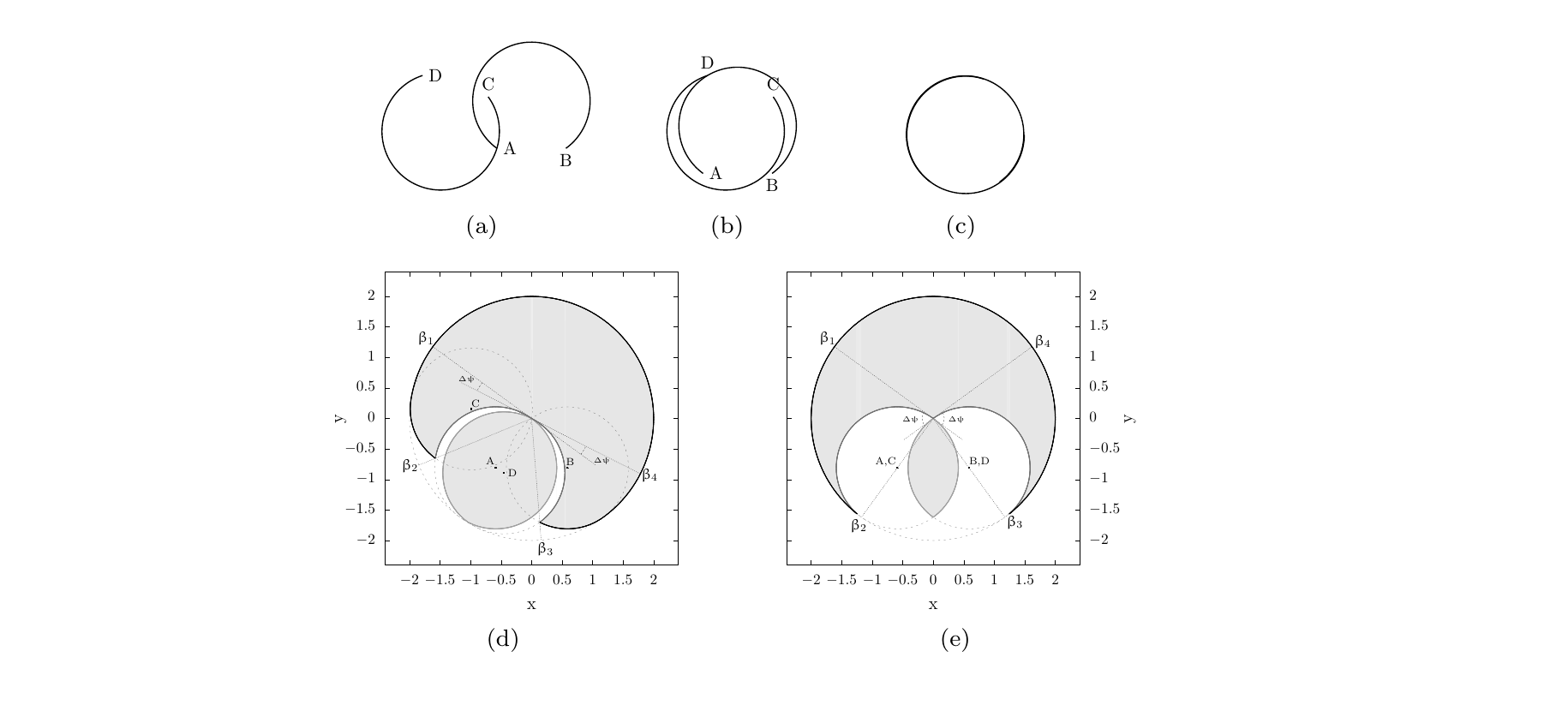}
\caption{
{\color{red}Examples of three} types of contact configuration between two hard major circular arcs in the case $\updelta < \uppsi \le \uppi$:
(a) the contact involves the concave side of one of the hard major circular arcs:
such contact pair configurations correspond to the dark gray sector in panels (d, e); 
(b) the two hard major circular arcs are intertwined: 
such contact pair configurations correspond to the light gray sector in panels (d, e); 
(c) the two hard major circular arcs are so closely intertwined that 
the centres of their parent circles essentially coincide:
such contact pair configurations correspond to the close neighbourhood of the point (0,0) in panels (d, e). 
In any of these panels, the extremal points of the hard major circular arc held fixed have been labelled as A and B while
those of the displacing hard major circular arc as C and D.
The excluded area, the {\color{red}shaded} region enclosed by 
the union of the black line, the dark gray line and the light gray line, of 
two hard circular arcs with  $\uptheta = 1.6 \uppi$, i.e $\updelta = 0.6 \uppi$,
for the angle of relative orientation $\uppsi = 0.65 \uppi$ (d) and $\uppsi = \uppi$ (e).  
Discontinuous round lines either correspond to the circumference with centre (0,0) and radius 2R or 
to circumferences with centres that are labelled as the extremal points in panels (a, b) and radius R.
The special angles are: 
$\displaystyle \upbeta_1 = \frac{\uppi}{2} + \frac{\updelta}{2}$,  
$\displaystyle \upbeta_2 = \frac{\uppi}{2} + \frac{\updelta}{2} + \frac{\uppsi}{2}$,
$\displaystyle \upbeta_3 = \frac{3\uppi}{2} - \frac{\updelta}{2} + \frac{\uppsi}{2}$ and 
$\displaystyle \upbeta_4 = \frac{3\uppi}{2} - \frac{\updelta}{2} + \uppsi$.
The angles indicated as $\Updelta \uppsi$ are those formed by the segments starting off the point (0,0)
and tangential to the dark gray and light gray lines; those segments that are tangential to the dark gray lines
correspond to the special angles $\upbeta_1$ and $\upbeta_4$; the amplitude of these angles is $\uppsi-\updelta$.  
Lengths in panels (d) and (e) are in units of R. 
}
\label{figura77}
\end{figure*}
In the case $\updelta < \uppsi \le \uppi$,
the characteristics of the excluded area significantly change.
In this case, 
three rather than two {\color{red}curve} sectors in 
the excluded--area boundary are recognised 
due to the notable insinuation of two no--overlap zones (Fig. \ref{figura77}).
The black external curve {\color{red}sector} in Fig. \ref{figura77} (d, e) corresponds to 
pair contact configurations of hard major circular arcs 
as arranged as in those pair contact configurations that 
have been mentioned in the case $0 \le \uppsi \le \updelta$ [Fig. \ref{figura33} (a, b)].
The dark gray intermediate curve {\color{red}sector} in Fig. \ref{figura77} (d, e) corresponds to 
pair contact configurations for which contact occurs on 
the concave side of one of the two hard major circular arcs [Fig. \ref{figura77} (a)].
The light gray internal curve {\color{red}sector} in Fig. \ref{figura77} (d, e) corresponds to
pair contact configurations for which contact occurs on 
the convex side of one of the two hard major circular arcs [Fig. \ref{figura77} (b)].
{\color{red}Notably,} the two horn--shaped regions delimited by the dark gray and light gray curves
correspond to non--overlapping pair configurations.
Differently than in the case $0 \le \uppsi \le \updelta$,
the two gray curves converge up to the point (0,0):
provided $\updelta < \uppsi \le \uppi$, 
the point (0,0) notably belongs to the excluded--area boundary [Fig. \ref{figura77} (d, e)].
The neighbourhoods of the point (0,0) within the two horn--shaped regions delimited
by the dark gray and light gray curves,
particularly those portions comprised between the two segments tangential
to these gray curves, with  the angle between these segments indicated as $\Updelta \uppsi $ [Fig. \ref{figura77} (d, e)],
correspond to non--overlapping pair configurations
for which two hard major circular arcs are specially intertwined [Fig. \ref{figura77} (b, c)].
{\color{red} By a succession of translatory and, possibly, also rotatory movements, 
that maintain} a value  of $\upalpha$ within $(\upbeta_1, \upbeta_2)$ or  within $(\upbeta_3, \upbeta_4)$, with
$\displaystyle \upbeta_1 = \frac{\uppi}{2} + \frac{\updelta}{2}$,  
$\displaystyle \upbeta_2 = \frac{\uppi}{2} + \frac{\updelta}{2} + \frac{\uppsi}{2}$,
$\displaystyle \upbeta_3 = \frac{3\uppi}{2} - \frac{\updelta}{2} + \frac{\uppsi}{2}$ and 
$\displaystyle \upbeta_4 = \frac{3\uppi}{2} - \frac{\updelta}{2} + \uppsi$,
two hard major circular arcs can come, without overlapping, 
so close that the centres of their parent circles effectively coincide [Fig. \ref{figura77} (c)]. 
The amplitude of these angular intervals $\Updelta \upbeta_{12}= \upbeta_2 - \upbeta_1$ and 
$\Updelta \upbeta_{34} = \upbeta_4 - \upbeta_3$ is proportional to 
the angle of relative orientation between the two hard major circular arcs: 
$\displaystyle \Updelta \upbeta_{12}$ = $ \displaystyle \Updelta \upbeta_{34} $ = $ \displaystyle \frac{\uppsi}{2}$.
Particularly, by maintaining a value of $\upalpha$ within the angular intervals indicated as $\Updelta \uppsi$
[Fig. \ref{figura77} (d, e)], 
once the two hard major circular arcs have come so close that the centres
of their parent circles effectively coincide, 
they can then stray around with the guarantee
that no overlap occurs between them.
The amplitude of $\Updelta \uppsi$ is equal to $\uppsi-\updelta$.
Equally as that of $\Updelta \upbeta_{12}$ and $\Updelta \upbeta_{34}$,
it linearly increases as the angle of relative orientation between the two hard major circular arcs increases.
Thus, on increasing the angle $\uppsi$, 
the two horn--shaped regions and, particularly, the portions associated to $ \Updelta \uppsi$  
widen [cf. panel (d) with panel (e) of Fig. \ref{figura77}] so that
specially intertwined pair configurations such as
those in Fig. \ref{figura77} (b, c) become increasingly more accessible. 
 
That the point (0,0) belongs to the excluded--area boundary {\color{red}can also occur} for 
hard minor circular arcs. 
For them, however, 
the point (0,0) belongs to this boundary 
if the value of $\uppsi$ is sufficiently small down to $\uppsi = 0$:
i.e., two hard minor circular arcs,  
as they are arranged on top of one another,
can arbitrarily closely approach.

This, seemingly minute, difference between 
the two sub--classes of hard circular arcs 
drastically affects the structure of their dense packings and the corresponding number density.
While for hard minor circular arcs there exist infinitely dense packings,
this is impossible for hard major circular arcs:
yet, {\color{red}very, albeit finitely, dense} packings of them can be analytically constructed exploiting 
their capability of intertwining, without overlapping, 
so closely that 
the centres of their parent circles essentially coincide.

\subsection{construction of the densest--known packings}
\label{condenspack}
That it is possible to closely intertwine two hard major circular arcs {\color{red}(section \ref{excar})} is exploited 
to demonstrate that 
it is actually possible to closely intertwine more
hard major circular arcs thus {\color{red} arriving at constructing} compact closed circular groups {\color{red}of them}. 
Specifically, it is demonstrated that, 
for a given angle $\updelta = \uptheta - \uppi$, 
one can arrange  $\displaystyle \mathsf{n} = \left [ \frac{2\uppi}{\updelta} \right ]$,  
with $\left [ x \right ] $ the strict floor function \cite{stretta},
hard major circular arcs so that the centres of their parent circles essentially coincide 
without making these hard particles overlap; 
i.e., $\mathsf{n}$ hard major circular arcs essentially arrange on the same circumference, 
thus maximally exploiting their concavity and infinitesimal thinness. 

First, one observes that is impossible 
to arrange more than $\mathsf{n}$ hard major circular arcs on the same circumference. 
If that were possible, 
there would exist a pair of hard major circular arcs whose 
angle of relative orientation $\uppsi$ would be smaller than $\updelta$. 
However, as shown in section \ref{excar},
two such hard major circular arcs would be incapable of 
approaching so closely to 
allow the centres of their parent circles to essentially coincide;
i.e., they would be incapable of arranging on the same circumference. 

Then, one demonstrates that 
it is exactly $\mathsf{n}$ the number of hard major circular arcs that 
can be arranged so that the centres of their parent circles essentially coincide{\color{red}; 
i.e.,
that can be arranged on the same circumference}.

\begin{figure}
\centering
\includegraphics[trim=0cm 2.5cm 0cm 1cm, clip, scale=0.2]{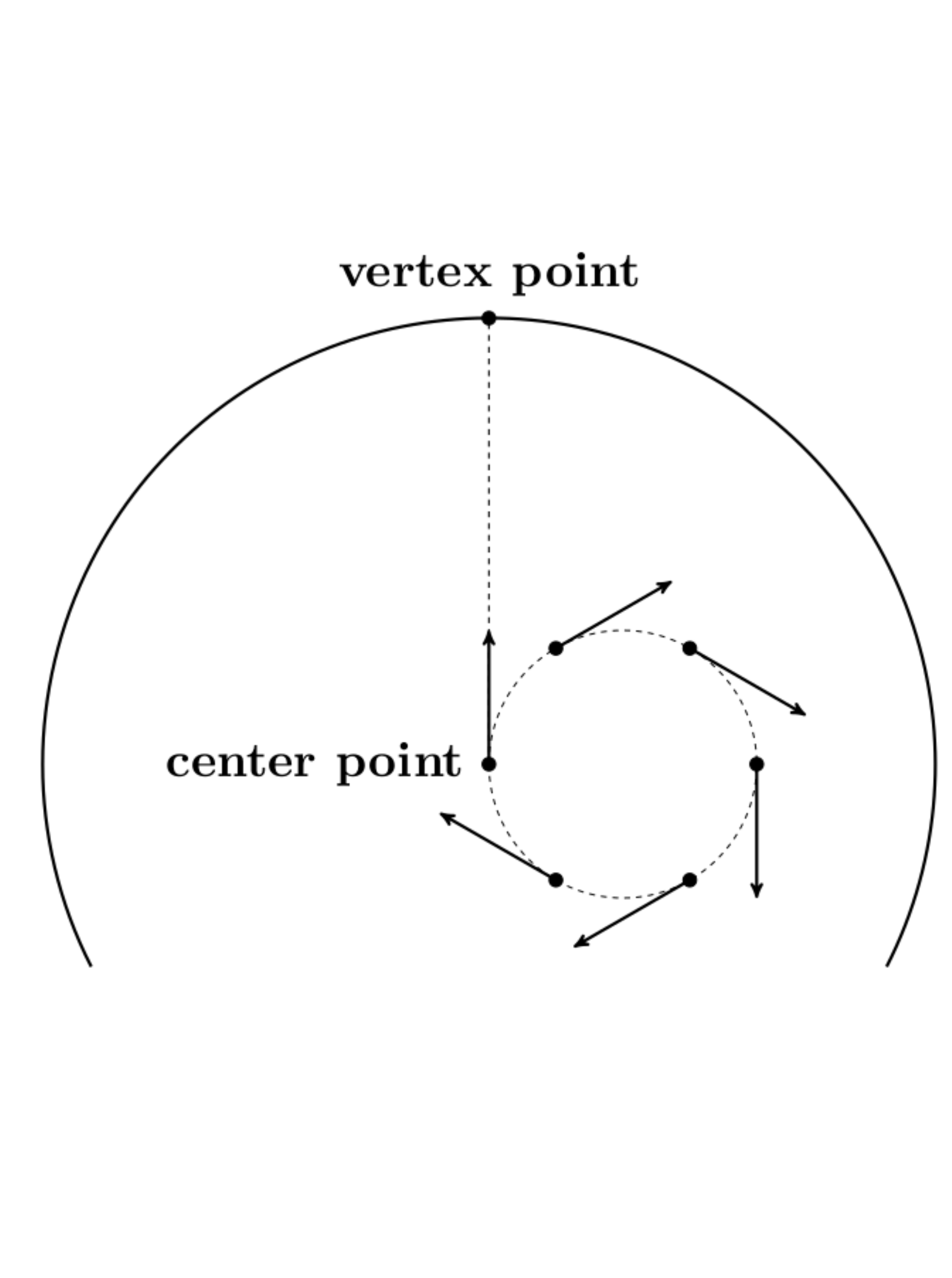}
\caption{For hard major circular arcs with $\uptheta = 1.3 \uppi$, i.e. $\updelta=0.3\uppi$, $\mathsf{n}=6$. 
These hard major circular arcs are arranged with 
the centres of their parent circles on the vertices of a regular {\color{red}convex} hexagon and 
their symmetry axes tangential to the circumscribing circumference 
{\color{red}whose radius is R$_{\rm crf}$ and (anti--)clockwise rotating.}
}
\label{arrangement}
\end{figure}

One begins {\color{red}by arranging}
the centres of the parent circles of these $\mathsf{n}$ hard major circular arcs at
the vertices of the corresponding regular {\color{red}convex} polygon whose 
characteristic {\color{red}(external)} angle is $\displaystyle \updelta^* = \frac{2\uppi}{\mathsf{n}}$.
Consider the circumference that circumscribes this polygon. 
The $\mathsf{n}$ hard major circular arcs are then arranged so that 
their symmetry axes are tangential to this circumference {\color{red}and (anti--)clockwise rotating}.
(Fig. \ref{arrangement} with the example of $\uptheta = 1.3 \uppi$, i.e. $\updelta=0.3\uppi$, for which $\mathsf{n}=6$).
Provided the radius of this circumference, R$_{\rm crf}$, 
is such that 
$\displaystyle \frac{{\rm R_{\rm  crf}}}{\rm R} < 
\frac{\sin \left ( \frac{\updelta^* - \updelta}{2} \right ) } { \sin  \left ( \frac{\updelta^*}{2} \right ) }$,
one now demonstrates that 
no two hard major circular arcs thus arranged can overlap.

\begin{figure}
\centering
\includegraphics[trim= 5cm 16.5cm 0cm 0cm,clip,scale=0.75]{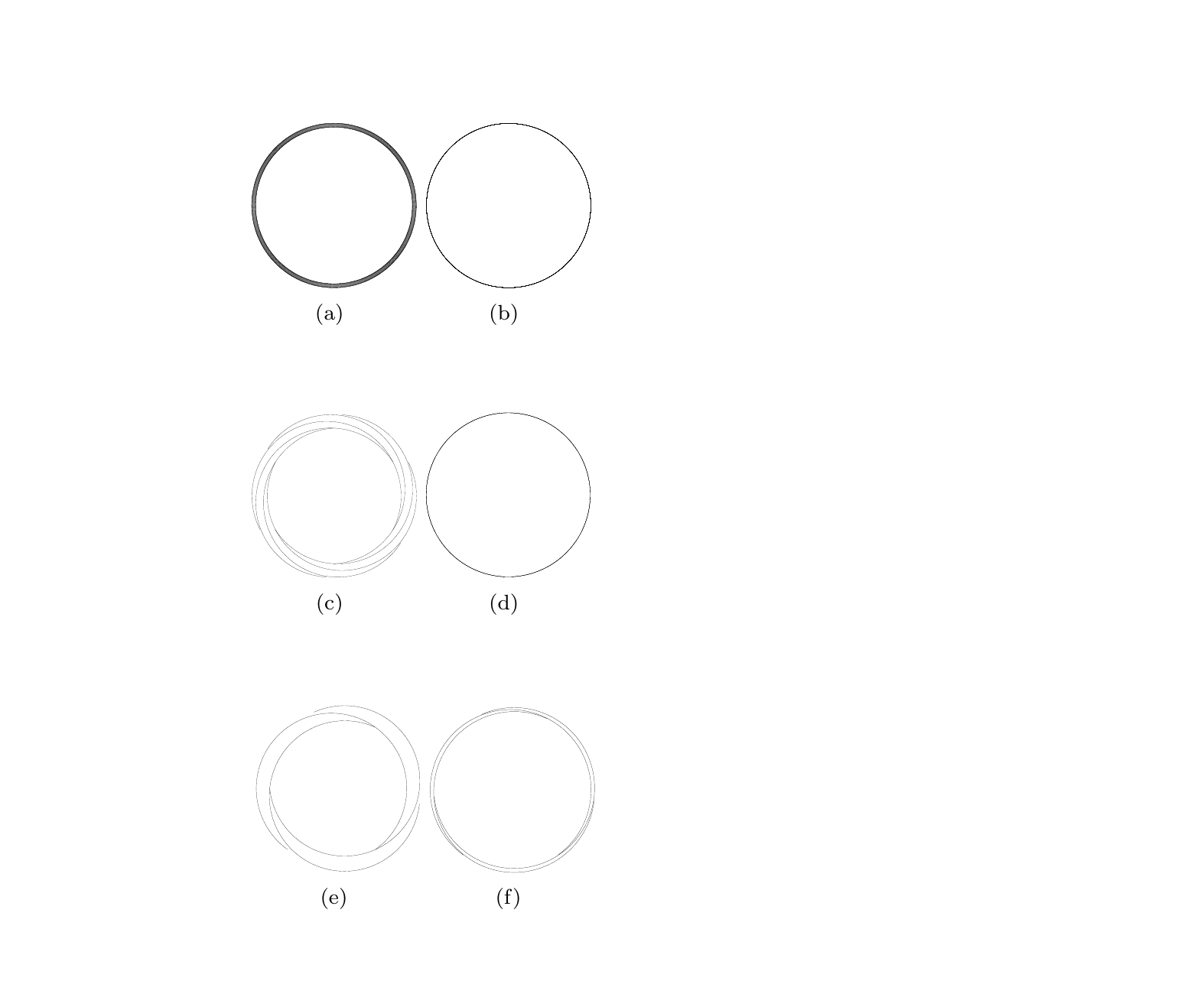}
\caption{
Examples of the compact closed circular groups that can be analytically constructed:
(a, b) $\uptheta = 1.05 \uppi$, i.e. $\updelta = 0.05 \uppi$, leading to 
$\mathsf{n} = 39$, and (a) ${\rm R}_{\rm cnf} = 0.025$ and (b) ${\rm R}_{\rm cnf} = 2 \times 10^{-12}$;
(c, d) $\uptheta = 1.3 \uppi$, i.e. $\updelta = 0.3 \uppi$, leading to 
$\mathsf{n} = 6$, and (c) ${\rm R}_{\rm cnf} = 0.1$ and (d) ${\rm R}_{\rm cnf} = 4 \times 10^{-5}$;
(e, f) $\uptheta = 1.6 \uppi$, i.e. $\updelta = 0.6 \uppi$, leading to 
$\mathsf{n} = 3$ and (e) ${\rm R}_{\rm cnf} = 0.11$ and (f) ${\rm R}_{\rm cnf} = 0.03$.
Lengths are in units of R. 
}
\label{multiplicerchi}
\end{figure}

By exploiting the regularity of the {\color{red}convex} polygon and the symmetry of the system, 
it suffices to ascertain the absence of an overlap between 
one hard major circular arc, taken as reference, 
with any one of the other  $\mathsf{n} - 1$ hard major circular arcs.
For a given reference hard major circular arc,
the angle of relative orientation of the $k$--th other hard major circular arc is $\uppsi_{k} = k\updelta^*$.
Since two relative orientations with angles $\uppsi_k$ and $\uppsi_{\mathsf{n} - k} = 2\uppi - k\updelta^*$ are equivalent,
it suffices to consider the angles of relative orientation $\uppsi_k \le \uppi$.
\begin{figure}
\centering
\includegraphics[trim=6.5cm 16.5cm 0cm 2cm, clip, scale=1]{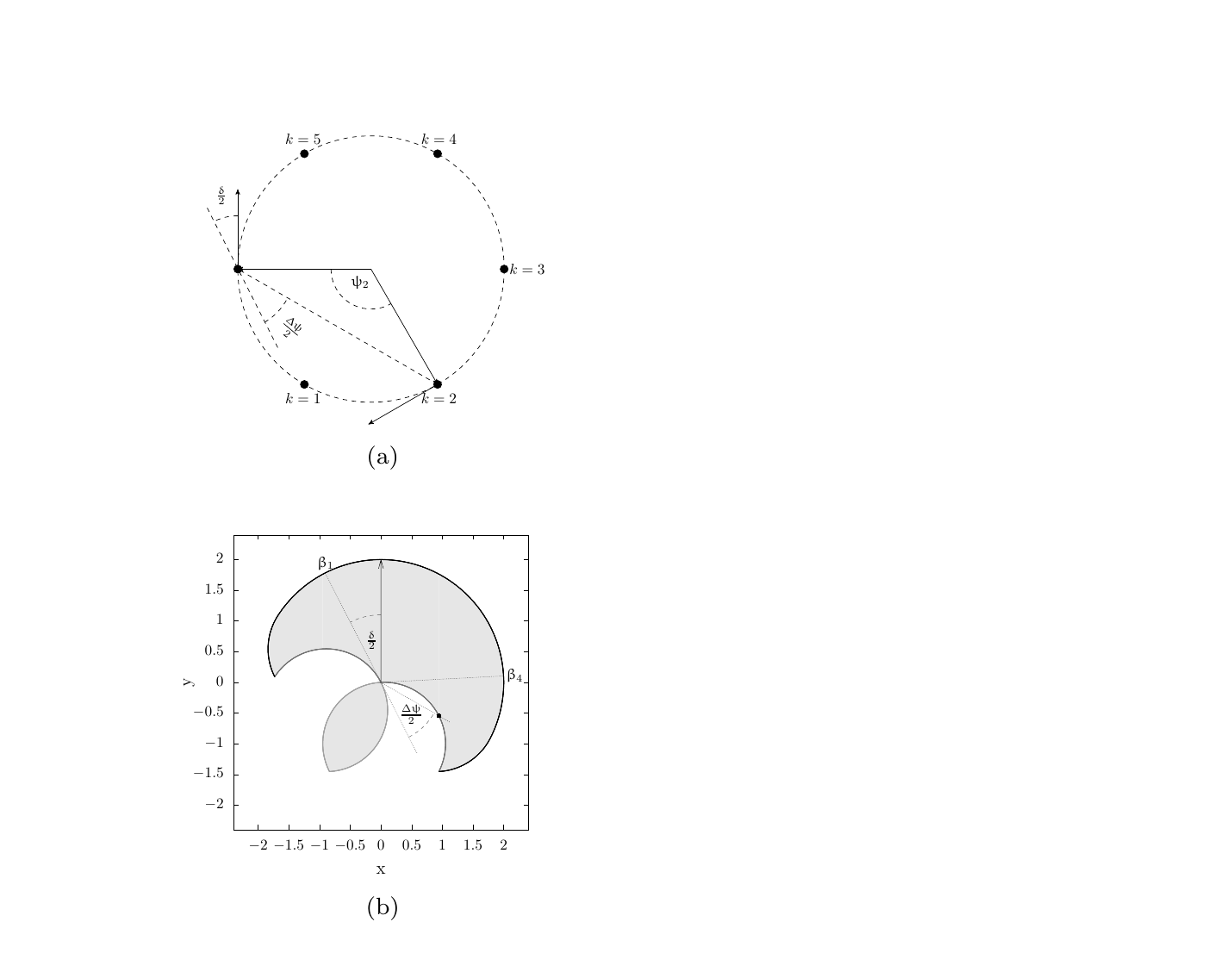}
\caption{(a) For a hard major circular arc with a subtended angle $\uptheta$ such that
$\mathsf{n}=6$, the centres of the parent circles are arranged on
the vertices of a regular {\color{red}convex} hexagon. Consider one hard major circular arc as a reference
with its symmetry axis aligned along the y--axis of the Cartesian reference frame and
the other $k=1, \dots, 5$ hard major circular arcs. 
Out of these, focus on $k=2$, the hard major circular arc whose symmetry axis forms an angle
$\uppsi_2$ with the symmetry axis of the reference hard major circular arc.
(b) The corresponding excluded area between the reference hard major circular arc and 
the $k=2$ hard major circular arc in (a). 
The black filled circle marks the intersection of 
the segment, 
that  starts off the point (0,0) and 
lies mid--way the segments tangential to the dark gray curve and light gray curve,
with this dark gray curve. 
The angles $\displaystyle \frac{\updelta}{2}$ and $\displaystyle \frac{\Updelta  \uppsi }{2}$ in (b) replicate those in (a). 
Lengths in panel (b) are in units of R.
}
\label{figura99}
\end{figure}
It is now useful to refer to the exemplificative and explicative  Fig. \ref{figura99} where,
in part (a),
the centres of the six parent circles of the hard major circular arcs 
are positioned at the vertices of a regular {\color{red}convex} hexagon while,
in part (b), 
the corresponding excluded area of two such hard major circular arcs,
whose relative orientation is $\uppsi_k$, is depicted for $k=2$.
If the reference hard major circular arc has its symmetry axis aligned along
the y--axis of the Cartesian reference frame, 
one can appreciate that
the angle $\upalpha$ of the distance vector that joins the centre of
its parent circle with the centre of the parent circle of the $k$--th other hard major circular arc 
lies mid--way the special angular interval comprised between the segments that
start off the point (0,0) and are tangential to the dark gray and light gray curves 
[Fig. \ref{figura77} (d, e) as well as Fig. \ref{figura99} (b)]. 
One {\color{red}recalls} that the amplitude of this angular interval is
$\Updelta \uppsi = \uppsi_k - \updelta$ 
{\color{red}(section \ref{excar})}.
If the angle $\upalpha$ is within such an angular interval, 
the $k$--th other hard major circular arc can arbitrarily closely approach 
the reference hard major circular arc.
The modulus of the distance that separates the centres of the parent circle of these two hard major circular arcs is equal to 
$\displaystyle 2{\rm R_{\rm  crf}} \sin \left (\frac{\uppsi_k}{2} \right)$.
This modulus has to be compared with 
the modulus of the distance between these two hard major circular arcs that 
corresponds to the intersection of the mid straight line with the dark gray curve [black filled circle in Fig. \ref{figura99} (b)].
One can appreciate that this modulus is equal to 2R$\sin \left (\frac{\uppsi_k-\updelta}{2} \right)$.
Provided ${\rm R_{\rm  crf}}$ is such that  
$\displaystyle \frac{{\rm R_{\rm  crf}}}{\rm R} < 
\frac{\sin \left ( \frac{k\updelta^* - \updelta}{2} \right ) } { \sin  \left ( \frac{k\updelta^*}{2} \right ) }$,
the $k$--th other hard major circular arc does not overlap with
the reference hard major circular arc.
One can observe that $\displaystyle \frac{{\rm R_{\rm  crf}}}{\rm R}$ is an increasing function of $k \updelta^*$.
Thus, it suffices that that condition is verified for $k=1$.
The condition 
$\displaystyle \frac{{\rm R_{\rm  crf}}}{\rm R} < 
\frac{\sin \left ( \frac{\updelta^* - \updelta}{2} \right ) } { \sin  \left ( \frac{\updelta^*}{2} \right ) }$
allows one to progressively shrink 
the circumscribing circumference on which the centres of the parent circles are positioned 
without this leading to the $\mathsf{n}$ hard major circular arcs overlapping.  
Thus, as R$_{\rm crf}$$\rightarrow 0$,  
{\color{red}the centres of the parent circles ultimately end up to essentially coincide,
i.e. the $\mathsf{n}$ hard major circular arcs} ultimately end up to 
essentially arrange on the same circumference (Fig. \ref{multiplicerchi}).

\begin{figure}
\centering
\includegraphics[trim=0.5cm 2.0cm 1.0cm 0.0cm,clip]{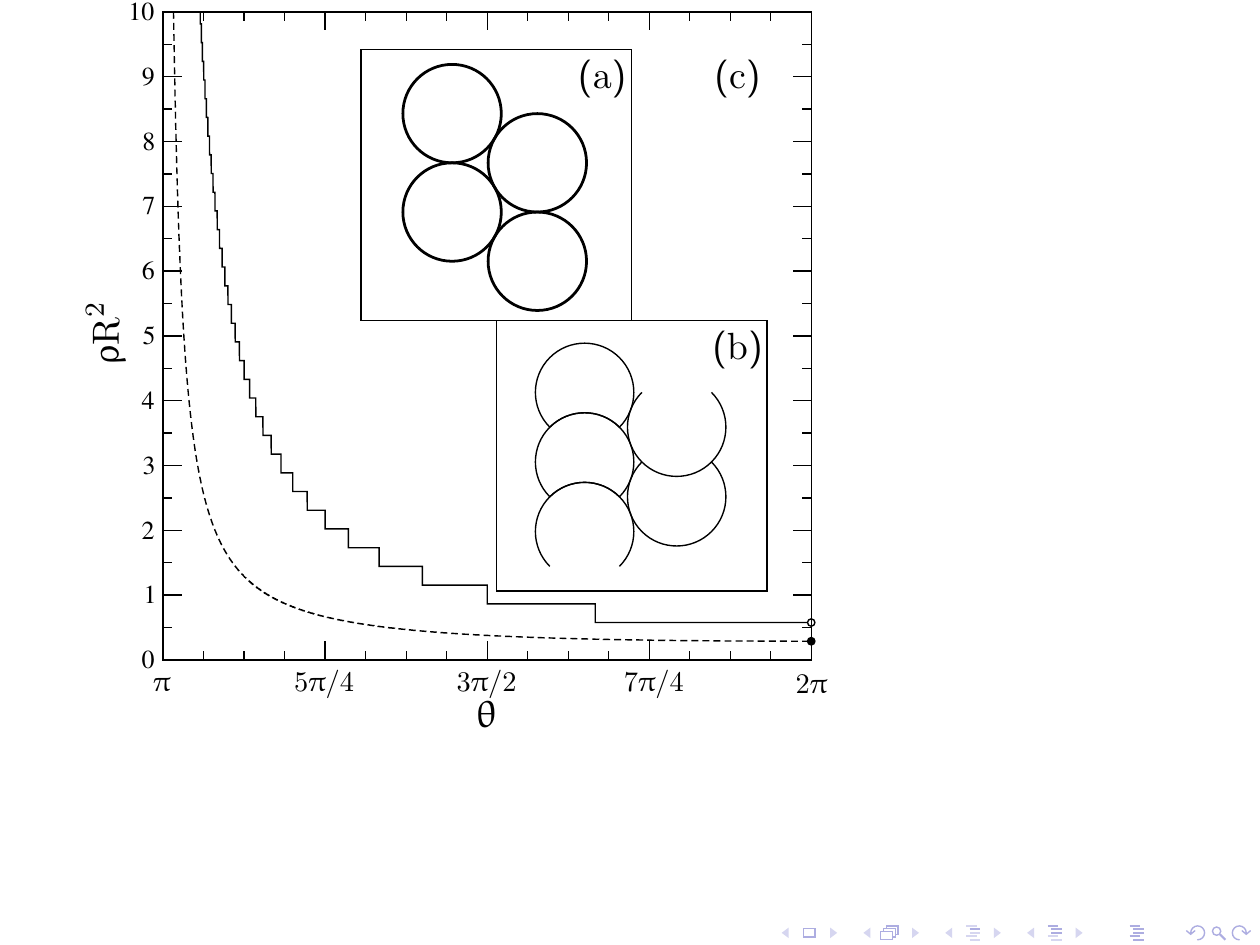}
\caption{
(a) Portion of a triangular lattice actually formed by 
the compact closed circular groups of $\mathsf{n}$ hard major circular arcs
as the thickness of the drawn circumferences wishes to indicate.
(b) Portion of a non--lattice formed by hard major circular arcs.
(c) Dimensionless number density, $\uprho {\rm R}^2$, as a function
of the subtended angle, $\uptheta$, for the generally non--lattice and non--periodic
compact--closed--circular--group triangular packings (black continuous step--like line) and 
the (double) lattice packings (black discontinous monotonic line) of hard major circular arcs.
Observe that, for $\uptheta =  2 \uppi$, 
$\uprho {\rm R}^2$ is equal to $\frac{1}{2\sqrt{3}}$ (black filled circle) rather than $\frac{1}{\sqrt{3}}$
(black empty circle) also for the former type of packings, 
{\color{red}in keeping with the known dimensionless number density of the densest packing} of hard circles
\cite{fejes}.  
}
\label{densepack}
\end{figure}

Having succeeded to construct compact closed circular groups of 
$\mathsf{n}$ hard major circular arcs,
it is then natural to arrange these circular groups on a triangular lattice [Fig. \ref{densepack} (a)].
These generally non--lattice and non--periodic infinitely degenerate \cite{explain} packings 
have a dimensionless number density  
\[ 
\displaystyle \uprho {\rm R}^2 \left(\uptheta \right) = 
\frac{1}{2\sqrt{3}} \left [ \frac{2\uppi}{\uptheta - \uppi} \right] = 
\frac{\mathsf{n}}{2\sqrt{3}} \,.
\]
Observe that the strictness of the floor function that defines $\mathsf{n}$ is necessary 
to recover the densest packing of hard circles and its number density \cite{fejes}  [Fig. \ref{densepack} (c)]. 
These generally non--lattice and non--periodic infinitely degenerate packings \cite{explain} constitute 
the densest--known packings for hard major circular arcs.  

One different packing strategy could {\color{red}have been} 
to arrange the hard major circular arcs in 
one of  the (non--)lattice infinitely degenerate packings where 
these hard particles are placed on top of one another as in Fig. \ref{densepack} (b) \cite{doppio}.
One can appreciate that the dimensionless number density of 
all these (non--)lattice packings \cite{doppio} is given by 
\[
\uprho {\rm R}^2  \left( \uptheta \right) =  \frac{1}{2 \cos \left ( \uppi - \frac{\uptheta}{2} \right) 
\sqrt{4- \cos^2\left( \frac{\uptheta}{2} \right) }} \,. 
\]
However, it is smaller than 
that of the packings formed by 
triangularly arranging 
the compact closed circular groups of $\mathsf{n}$ hard major circular arcs
[Fig. \ref{densepack} (c)].

One {\color{red}could have attempted} to combine the two packing strategies. 
First, one essentially arranges a number $m$ of 
hard major circular arcs on 
the same circumference by 
successively rotating them of an angle infinitesimally larger than $\updelta$. 
The number $m = 1, \dots, \mathsf{p}$, 
with $ \displaystyle {\mathsf{p}}=\left [ \frac{\uppi}{\updelta} \right]$ \cite{stretta},
is sufficiently small
that the group thus constructed is generally arched and open
with an effective subtended angle equal to 
$\uptheta^* = \uptheta + (m-1) \updelta = \uppi + m\updelta$.
Then, one arranges these arched open groups as 
it has been done with a single hard major circular arc in Fig. \ref{densepack} (b).
The dimensionless number density of such packings is:
\[
\uprho {\rm R}^2  \left( \uptheta \right) =  \frac{m}{2 \cos \left ( \uppi - \frac{\uptheta^*}{2} \right) 
\sqrt{4- \cos^2\left( \frac{\uptheta^*}{2} \right) }} \, .
\]  
Irrespective of the value of $\updelta$, 
this function monotonically increases with $m$, 
attaining its maximum at $m = {\mathsf{p}}$, i.e.
once the arched and open groups have actually become closed and circular and 
ended up to arranging  on a triangular lattice.
However, $\mathsf{p} < \mathsf{n}$: 
the circular and closed groups thus constructed 
are not as numerous as those constructed by 
the method  based on 
positioning the centres of their parent circles at the vertices of a regular {\color{red}convex} polygon:
consequently, the dimensionless number density achieved by 
these packings is smaller than that of 
the packings formed by triangularly arranging the compact closed circular
groups of $\mathsf{n}$ hard major circular arcs.

\section{Monte Carlo calculations}
\label{MC}

One may inquire whether 
closed circular groups similar to those 
{\color{red}described in section \ref{condenspack}} that 
constitute the structural units of 
the densest--known packings of hard major circular arcs
may spontaneously form and 
how compact and numerous they result to be.

To {\color{red}address this point}, 
specific Monte Carlo (MC) \cite{MCorigin,MCNPT,MCallen} calculations were carried out.
In these calculations,
systems of $N$ hard major circular arcs in 
a deformable parallelogrammatic container with hard walls
were compressed from a low to a high pressure.
The initial value of $N$ was 2.
These two hard major circular arcs were initially placed in 
a large square container with hard walls.  
The initial value for 
the dimensionless pressure 
$ \displaystyle P^* = \frac{P {\upsigma}^2}{k_B T} $, 
with $P$ the pressure, 
$\upsigma^2$ the area of 
that part of a spherical surface subtended by 
the angle $\uptheta$, 
$k_B$ the Boltzmann constant and 
$T$ the absolute temperature, was 1.
Successively, $P^*$ was gradually increased in 
logarithmic steps up to a value equal to 100. 
For each of these successive values of $P^*$, 
1 million of MC cycles were usually carried out, 
with a MC cycle defined as 
a set of $N$ random translations of a randomly selected particle,  
$N$ random rotations of a randomly selected particle and 
one random change of the length and/or orientation of  
a randomly selected side of the container.
In the course of these MC calculations,
the maximal sizes for a random translation, 
a random rotation and a random change  of 
a side of the container
were progressively adjusted to ensure that 
a fraction comprised between 0.2 and 0.3 of 
{\color{red}each of} these trial moves was accepted.
The acceptance of a new shape and a new size for the container was subject to 
the usual Metropolis criterion of a constant--pressure MC calculation \cite{MCNPT,MCallen}.
On completion of this sequence of MC calculations, 
it was observed whether a closed circular group was formed.
The number of hard major circular arcs $N$ was then increased by one and 
the same sequence of MC calculations was repeated for that new value of $N$.
This incremental addition of hard major circular arcs and 
successive MC--method--based compression of the system were carried on until 
it was observed that the formation {\color{red} of}
a closed circular group of $N$ hard major circular arcs did not occur.  
The largest value of $N$ that led to 
the formation of a closed circular group 
was then registered.
For a number of values of $\uptheta$, 
this entire process was repeated a number of times, usually seven,
changing the initial configurations of the $N$ hard major circular arcs and 
the seed of the random number generator \texttt{mt19937} \cite{mt19937}.
By operating in this way, 
Table \ref{tabella} was constructed. 
For a number of values of $\uptheta$,
it reports the seven largest values of $N$ that 
led to the formation of a closed circular group. 
These numerical results prove that 
the spontaneous formation of closed circular groups is indeed possible, 
even for a value of $\uptheta$ as large as $359/180 \uppi$.   
Yet, one has to also observe that 
these moderately fluctuating largest values of $N$ are 
usually incapable of equalling the 
larger number $\mathsf{n}$ of hard major circular arcs  that 
is possible to closely intertwine by analytic construction (section \ref{condenspack}).

\begin{table}
\begin{tabular}{|c||c|c|c|c|c|c|c||c|}
\hline
$\uptheta$ & $N_1$ & $N_2$ & $N_3$ & $N_4$ & $N_5$ & $N_6$ & $N_7$ & $\mathsf{n}$ \\
\hline
\hline
$\uppi$      & 16 & 17 &  17 &  17 &  17 & 16 & 16 & $\infty$\\
\hline
$1.05 \uppi$ & 13 & 11 &  11 &  11 &  11 & 12 & 12 & 39 \\
\hline
$1.1 \uppi$  &  8 &  8 &   9 &   9  & 9 &  9 &  9  & 19 \\
\hline
$1.15 \uppi$  & 8 & 7 & 8 & 7 & 7 & 7 & 7 & 13 \\ 
\hline
$1.2 \uppi$   & 6 & 6 & 6 & 6  & 7 & 6 & 6 & 9  \\
\hline
$1.25 \uppi$  &  6  & 5 &  5  & 5 & 6  & 5 & 5 & 7  \\ 
\hline
$1.3 \uppi$   &  5 & 5 & 5 & 4 & 4 & 5 &  5 &  6 \\
\hline
$121/90 \uppi$ &  4  & 4  & 4  &  4 & 4 & 4 & 4  &  5 \\
\hline
$1.4 \uppi$  &  4 & 4 & 4 & 4 & 4 & 4 & 4  &  4 \\
\hline
$1.45 \uppi$ &  3 & 3 & 4 & 3 & 3 & 3 & 3 &  4 \\ 
\hline
$1.5 \uppi$  &  3 & 3 & 3 & 3 & 2 & 3 & 3 &  3 \\
\hline
$1.55 \uppi$ &  2 & 2 & 3 & 2 & 2 & 2 & 2 &  3 \\  
\hline
$1.6 \uppi$  &  2 & 2 & 2 & 2 & 2 & 2 & 2  &  3 \\
\hline
$1.95 \uppi$ &  2 & 2 & 2 & 2 & 2 & 2 & 2  &  2 \\
\hline
$359/180 \uppi$  & 2 & 2 & 1 & 2 & 2 & 1 & 1  & 2 \\
\hline
\end{tabular}
\caption{Largest value of the number of 
hard major circular arcs per  closed circular group, $N_i$, 
attained in the $i$--th sequence of Monte Carlo calculations on compression, 
as well as the corresponding $\mathsf{n}$,  
as a function of the subtended angle $\uptheta$.}
\label{tabella}
\end{table}

Based on these results,
one may then inquire whether 
the compact closed circular groups
that are analytically constructed can 
spontaneously unfasten when a
system of them is decompressed in 
analogous MC calculations.
For values of the angle $\uptheta \ge 1.2 \uppi$,
starting from a dimensionless pressure 
equal to 100 and decreasing it by logarithmic steps
until a value of 0.01 was reached, 
it proved relatively easy 
to completely unfasten the analytically constructed compact close circular groups.
For values of the angle $\uptheta < 1.2 \uppi$, 
this procedure was insufficient:
these more compact and numerous analytically constructed closed circular groups
survived down to that very small value of $P^*$.
Nevertheless, 
they ultimately succeeded to unfasten in relatively painful MC calculations that 
protracted up to $10^3$ million of MC cycles once they were
suitably modified so that the random change of 
the container shape and size was accepted only if 
it led to a larger container area.  

\section{conclusion}
\label{conclusion}
In this work, densest--known packings of congruent hard infinitesimally--thin circular arcs are analytically constructed.
The interest is actually in 
those hard infinitesimally--thin circular arcs denotable as major whose subtended angle $\uptheta \in (\uppi, 2\uppi]$.
In spite of being infinitesimally thin, 
there exists no pair configuration of them where they arbitrarily closely approach and which
is replicable ad infinitum, without introducing any overlap,
as it occurs for hard infinitesimally--thin minor circular arcs
whose subtended angle  $\uptheta \in [0,\uppi]$. 
Nevertheless, it is shown that hard infinitesimally--thin major circular arcs can be carefully arranged in 
compact closed circular groups formed by 
$\displaystyle \mathsf{n} = \left [ \frac{2\uppi}{\uptheta - \uppi} \right ]$ 
\cite{stretta} of them. 
These compact closed circular groups can then be arranged with 
the communal centres at the sites of a triangular lattice thus 
leading to generally non--lattice non--periodic infinitely degenerate packings with 
dimensionless number density equal to $\displaystyle \frac{\mathsf{n}}{2\sqrt{3}}$.
These densest--known packings are classifiable as purely entropy--driven cluster {\color{red}(porous)} crystals.
It is notable that very simple particles, 
such as these hard infinitesimally--thin concave particles,
devoid of any attractive or complicated interactions between them, 
can first cluster and 
then these clusters act as structural units to form a {\color{red}(porous)} crystal.
Monte Carlo calculations confirm that these clusters,
albeit not as numerous as those analytically constructed, 
can spontaneously form on compression.

On these premises, 
the investigation of 
the complete phase diagram of systems of 
hard infinitesimally--thin circular arcs
will prove interesting.
{\color{red}
Once the latter phase diagram will have been mapped,
one will possibly move on 
to considering dense packings and phase behaviour of 
hard finitely--thin circular arcs.
In actuality, one real example of these systems has been recently investigated \cite{jacs,prejacs}: 
hard colloidal \textsf{C}--shaped particles,
corresponding to hard finitely--thin circular arcs with $\uptheta=\displaystyle\frac{3}{2}\uppi$,
have been prepared and their phase behaviour 
first experimentally investigated in a tilted two--dimensional gravitational column \cite{jacs} and 
then theoretically rationalised \cite{prejacs}:
dimerisation has been observed.
It is presumable that progressive hard--(colloidal, granular)--particle thickening will increasingly impede reaching 
a number of intertwined hard (colloidal, granular) major circular arcs as large as in the infinitesimally--thin case but 
it should still prove interesting to generally examine up to which extent.}


\begin{thebibliography}{99, numbers, square}
\bibitem{stretta}
The strict floor function is $\left[ x \right] = - 1 - \lceil -x \rceil$, 
with $\lceil x \rceil$ the ordinary ceiling function.
\bibitem{explain}
These packings are generally non--lattice \cite{torquato1,torquato2} as their fundamental cell generally contains
more than one hard particle.
They are also generally non--periodic \cite{torquato1,torquato2} as
the hard major circular arcs that form a compact closed circular group can be arranged either
anti-- or clockwise and 
the compact closed circular groups can be rotated of an arbitrary angle with respect to one another. 
They are infinitely degenerate as these arbitrary rotations do not affect the value of the number density.
\bibitem{doppio}
Irrespective as to whether the hard major circular arcs orient 
the unit vectors that join the centre of their parent circle with the respective vertex 
either parallel, i.e. they form a lattice packing,
or anti--parallel, i.e. they form a non--lattice packing such as the one in Fig. \ref{densepack} (b),
the number density is the same.
Each of these (non--)lattice packings of hard major circular arcs coincides with 
the corresponding packing of the corresponding hard lunate particles such as in Fig. \ref{figura55}.
\end{thebibliography}
\end{document}